\documentclass[sigconf]{acmart}
\usepackage{multirow}

\AtBeginDocument{%
  \providecommand\BibTeX{{%
    \normalfont B\kern-0.5em{\scshape i\kern-0.25em b}\kern-0.8em\TeX}}}

\copyrightyear{2021}
\acmYear{2021}
\setcopyright{acmcopyright}\acmConference[ICMI '21]{Proceedings of the 2021 International Conference on Multimodal Interaction}{October 18--22, 2021}{Montréal, QC, Canada}
\acmBooktitle{Proceedings of the 2021 International Conference on Multimodal Interaction (ICMI '21), October 18--22, 2021, Montréal, QC, Canada}
\acmPrice{15.00}
\acmDOI{10.1145/3462244.3479967}
\acmISBN{978-1-4503-8481-0/21/10}

\begin{document}

\title{Multimodal Approach for Assessing Neuromotor Coordination in Schizophrenia Using Convolutional Neural Networks}

\author{Yashish M. Siriwardena}
\email{yashish@umd.edu}
\authornotemark[1]
\author{Carol Espy-Wilson}
\email{espy@umd.edu}
\affiliation{%
  \institution{University of Maryland}
  \city{College park}
  \state{MD}
  \country{USA}
}

\author{Chris Kitchen}
\email{Ckitchen@jhu.edu}
\author{Deanna L. Kelly}
\email{dlkelly@som.umaryland.edu}
\affiliation{%
  \institution{University of Maryland School of Medicine}
  \city{Baltimore}
  \state{MD}
  \country{USA}
}

\begin{abstract}
    This study investigates the speech articulatory coordination in schizophrenia subjects exhibiting strong positive symptoms (e.g. hallucinations and delusions), using two distinct channel-delay correlation methods. We show that the schizophrenic subjects with strong positive symptoms and who are markedly ill pose complex articulatory coordination pattern in facial and speech gestures than what is observed in healthy subjects. This distinction in speech coordination pattern is used to train a multimodal convolutional neural network (CNN) which uses video and audio data during speech to distinguish schizophrenic patients with strong positive symptoms from healthy subjects. We also show that the vocal tract variables (TVs) which correspond to place of articulation and glottal source outperform the Mel-frequency Cepstral Coefficients (MFCCs) when fused with Facial Action Units (FAUs) in the proposed multimodal network. For the clinical dataset we collected, our best performing multimodal network improves the mean F1 score for detecting schizophrenia by around 18\% with respect to the full vocal tract coordination (FVTC) baseline method implemented with fusing FAUs and MFCCs.
\end{abstract}



\maketitle

\section{Introduction}
Schizophrenia is a chronic mental disorder with heterogeneous
presentations that affect around 60 million (1\%) of the world’s adult population \cite{kuperberg}. Symptoms of schizophrenia are broadly categorized as either positive, which are pathological functions not present in healthy individuals (e.g., hallucinations and delusions); negative, which involve the loss of functions or abilities (e.g., apathy, lack of pleasure, blunted affect and poor thinking); or cognitive (deficits in attention, memory and executive functioning) \cite{Andreason, demily}. 
Previous studies have shown promising results in identifying the severity of depression by using coordination features derived from the correlation structure of the movements of various articulators \cite{Espy-Wilson2019}. Based on this, a preliminary study was done by Siriwardena et al. \cite{Yashish2020} to understand how positive symptoms of schizophrenia affect the articulatory coordination in speech. These findings are the impetus for the current investigation where more subjects and data are used to validate the fact that neuromotor coordination is altered in schizophrenic patients who are markedly ill and exhibit strong positive symptoms.  


Time-delay embedded correlation (TDEC) analysis  has shown promising results in assessing neuromotor coordination in Major Depressive Disorder (MDD), and the eigenspectra derived from the correlation matrices have been used effectively for classification of MDD subjects from healthy \cite{Williamson2014, WILLIAMSON2019, Seneviratne2020}. Recently, new multi-scale full vocal tract coordination (FVTC) features generated with a dilated CNN have shown further improvement in classification for selected datasets of MDD subjects \cite{Huang2020ExploitingVT}. The FVTC method addresses repetitive sampling and matrix discontinuity issues of TDEC analysis by introducing a new channel-delay correlation matrix. In this paper, we compare both TDEC and FVTC methods for generating input correlation matrices for training a multimodal CNN with audio and video features. We also propose a model which uses both TDEC and FVTC correlation structures to classify subjects with strong positive symptoms in schizophrenia from healthy. 


Throughout the paper, we compare and contrast studies mostly done in depression analysis since it is hard to find any comparable speech articlulatory coordination based analysis on schizophrenia. While there are studies investigating the changes in language used when subjects are schizophrenic with positive symptoms \cite{MOMENI2012288}, to our knowledge this study is one of the first papers to present results that show changes in the coordination of speech gestures produced by schizophrenic subjects. Extending the results in \cite{Yashish2020}, this paper also validates that schizophrenic subjects with strong positive symptoms have a more complex articulatory coordination with respect to healthy controls. Finally, this study presents the importance and effectiveness of multimodal fusion (audio and video) for screening mental health disorders like schizophrenia by proposing a new CNN based multimodal architecture. 

\section{Database and Features}
\subsection{Database}
\label{ssec:subhead}

A database recently collected for a collaborative observational study conducted by the University of Maryland School of Medicine and the University of Maryland College Park has been used for this study \cite{UMBdataset}. The database contains video and audio data of free response assessments administered in an interview format. Data for this study was collected from 23 schizophrenic (SZ) patients and 20 healthy controls (HC). All of the schizophrenic patients were clinically diagnosed. Every subject participated in four interview sessions over a period of six weeks. Each interview session is 10-45 minutes long and every subject is assessed using standard depression severity measures and global psychopathology measures by a clinician and themselves. For this study, we used the clinician assessments based on the 18-item Brief Psychiatric Rating Scale (BPRS) \cite{BPRS_Hunter2011}, where we selected subjects based on the total BPRS score, and the subscores for psychosis (BPRS item11, item12, item4, item15) and activation (BPRS item6, item7, item17), and the Hamilton Rating Scale for Depression (HAMD) \cite{HAMD_Gonzalez2013}.
 
Table \ref{tab:data_records} presents the information on the dataset used for the study. The 7 schizophrenic subjects (4 Males, 3 Females) are selected such that they are markedly ill (BPRS total $\geq$ 45), have higher sub scores for psychosis and activation, but are not depressed or only mildly depressed (HAMD between 0 and 14). The 11 healthy controls (5 Males, 6 Females) are chosen such that they are not depressed (HAMD $<$ 7) or schizophrenic (BPRS $<$ 32). 
 
The audio data were first diarized using transcripts which include the speaker ID and time stamps to separate out the utterances which correspond to the subject from the interviewer. The utterances which are longer than 40 seconds were then segmented into 40 second chunks.  If the remaining amount was less than 5 seconds (the minimum length accepted), then it was added back to the last segment. Thus, for all the classification experiments, we used segments with a minimum length of 5 seconds and a maximum length of 45 seconds.

\begin{table}[t]
  \caption{Details of the dataset used}
  \centering
  \label{tab:data_records}
  \begin{tabular}{cp{7mm}p{8.5mm}p{8mm}p{7mm}c}
    \toprule
    \textbf{}
    & \textbf{SZ} 
    & \textbf{HC}   \\
    \midrule
Number of subjects                  & 7      & 11\\  
Number of sessions                  & 17     & 34 \\  
Mean session duration           & 35min & 18min \\
Number of utterances                & 1208   & 1132\\   
Hours of speech                 & 10.0     & 9.43\\  
    \bottomrule
  \end{tabular}
 \label{tab:dataset}
\end{table}

\subsection{Facial Action Units (FAUs)}
\label{ssec:fauextract}

The video-based Facial Action Units (FAUs) provide a formalized method for identifying changes in facial expressions. We used the Openface 2.0: Facial Behaviour Analysis toolkit \cite{openface} to extract seventeen FAUs from the recorded videos of the subjects during the interviews. The FAU features were sampled at a rate of 28 frames per second.  We only analyzed those portions of the video when the subject was talking. 
For parallel delay CNN model in section \ref{ssec:paralleldelayCNN} and FVTC model in section \ref{ssec:FVTC}, we choose only 10 FAUs (FAU 6,7,9,10,12, 14,15,17,20 and 23 from FACS coding system \cite{Prince2015FacialAC}) which are near the mouth area that can capture coordination of lip and near lip movements during the voice activity. One other reason to choose 10 FAUs is to handle the high dimensionality of the TDEC correlations structure which poses computational limitation when training multi-modal networks in section \ref{ssec:multimodalCNN}.

    
  
\subsection{Vocal Tract Variables (TVs)}

We used a speech inversion (SI) system \cite{Sivaraman2016, Sivaraman2017a} that maps the acoustic signal into 6 vocal tract variables (TVs). The 6 TVs are namely Lip Aperture (LA), Lip Protrusion (LP), Tongue Body Constriction Location (TBCL), Tongue Body Constriction Degree (TBCD), Tongue Tip Constriction Location (TTCL) and, Tongue Tip Constriction Degree (TTCD).

Seneviratne et al. \cite{Seneviratne2020} in a recent study showed that incorporating glottal TVs generated by periodicity and aperidocity measures (by Aperiodicity, Periodicity and Pitch (APP) detector \cite{APPdetector}) improved the results of depression detection. Thus, in this study we use 6 TVs generated from the SI along with the 2 glottal TVs as the key audio features for the classification models.

\subsection{Mel-Frequency Cepstral Coefficients (MFCCs)}

Previous studies in depression prediction using speech \cite{avec_winning1, 2019avec} have shown the superiority of MFCCs over other audio based features like extended Geneva Minimalistic Acoustic Parameter Set (eGeMAPS) \cite{eGeMAPS} and DEEP SPECTRUM features \cite{deepSpectrumfeats}. Huang et al. \cite{Huang2020ExploitingVT} showed with their depression classification study that coordination features computed from MFCCs perform better with respect to formants and eGeMAPS features. So to compare how robust and effective the TVs are for detecting schizophrenia, we chose MFCCs as the baseline audio features for our study. We extracted 13 MFCCs from the librosa python library using an analysis window of 20 ms with a 10 ms frame shift. Only 12 MFFCs were used for analysis by discarding the 1st coefficient. 

\section{Methodology}
\label{sec:methodology}
\subsection{Time-delay embedded correlation Analysis (TDEC)}
\label{ssec:paralleldelayCNN}

Coordination among 10 FAUs, 6 TVs and the 12 MFCCs were estimated using the correlation structure features. The features are estimated by computing a channel delay correlation matrix using time delay embedding at two delay scales \cite{Espy-Wilson2019, WILLIAMSON2019, Williamson2013}. The computed correlation matrix can be considered as a compact representation which provides rich information on the underlying mechanisms in articulatory coordination. Each correlation matrix $R_{i}$ has a dimensionality (MN x MN) where M = 10, 8 and 12 for FAUs, TVs and MFCCs respectively. N corresponds to the number of time delays per channel which is 15 for all the considered feature types.

From the correlation matrix $R_{i}$ calculated for each sample $i$, the eigenspectrum is computed. The eigenspectrum generated for FAUs is a 150-dimensional vector which is rank ordered (in descending order of magnitude of eigenvalues) from index j=1,..,150. The eigenspectra generated from TVs and MFCCs are 120-dimensional and 180-dimensional, respectively.

\begin{figure}
	\begin{minipage}[0.49\columnwidth]{
	   0.23\textwidth}
	   \includegraphics[width=\linewidth, scale=0.9, height=9.5cm]{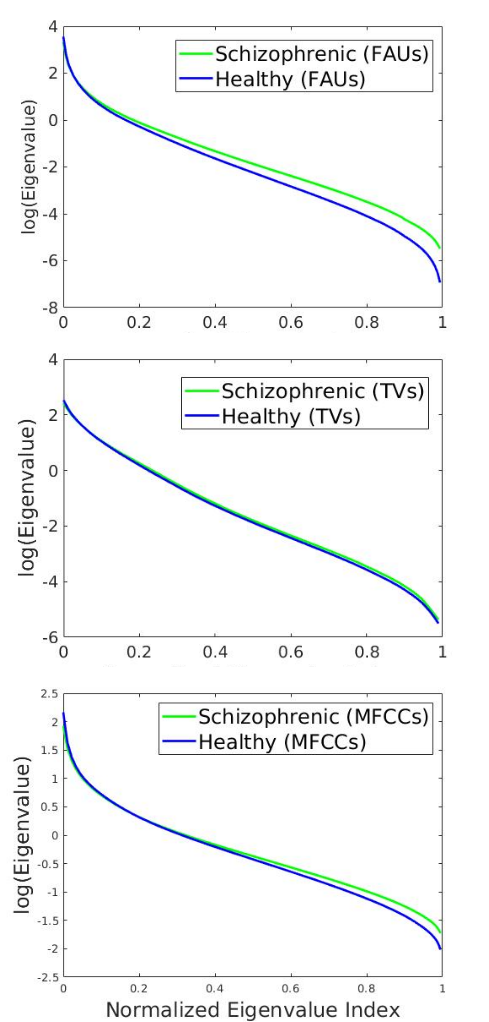}
	   \label{fig:eigenspecs}
	\end{minipage}
	\begin{minipage}[0.49\columnwidth]{
	   0.23\textwidth}
	   \includegraphics[width=\linewidth, scale=0.9, height=9.5cm]{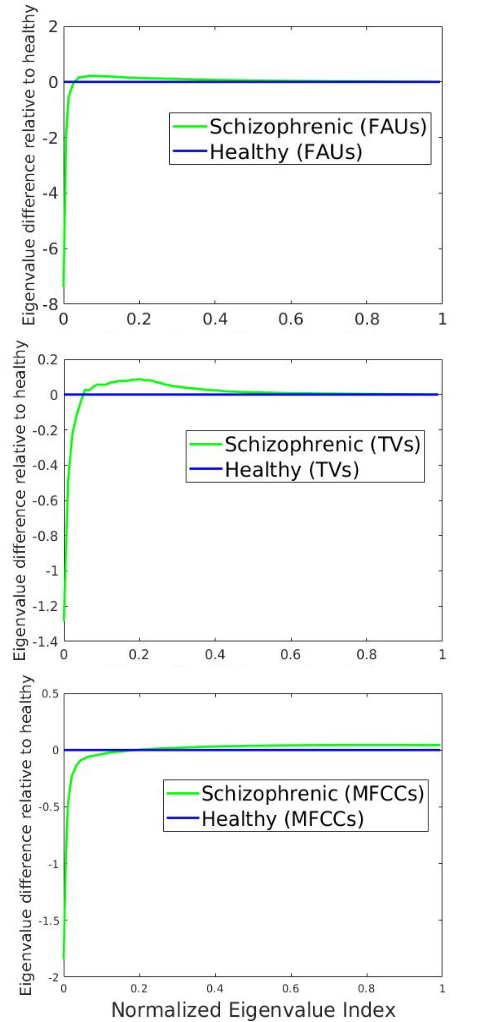}
	   \label{fig:cohend}
	\end{minipage}
	\caption{Averaged eigenspectra (on left) and the corresponding difference plots (on right) for FAUs, MFCCs and TVs}
    \label{fig:eigenspecs and cohends}
\end{figure}



Figure \ref{fig:eigenspecs and cohends} shows the averaged eigenspectra (on left) computed for FAUs, TVs and MFCCs. The difference plots in Figure \ref{fig:eigenspecs and cohends} (in right) are calculated by taking the difference between averaged eigenspectra curve for speech from schizophrenic subjects with respect to that of healthy controls. These eigenspectra and difference plots help us to understand the coordination complexity of the speech and facial gestures. The magnitude of each eigenvalue is proportional to the amount of correlation in the direction of their associated eigenvectors. The difference plots show that schizophrenic speech has smaller low-rank eigenvalues relative to the healthy controls, and the trend is reversed towards the high-rank eigenvalues. Therefore, schizophrenic speech needs a larger number of independent dimensions implying a more complex articulatory coordination than speech from healthy controls \cite{Williamson2013, EEG_WILLIAMSON}. The same argument is true for facial gestures.


\subsection{Full vocal tract coordination (FVTC)}
\label{ssec:FVTC}

Huang et al. \cite{Huang2020ExploitingVT} in a recent study with MDD introduces a new channel delay correlation method inspired by TDEC, which uses a different correlation structure with correlations starting from 0 to a delay of 'D' frames (a design choice). The delayed autocorrelations and cross-correlations across channels are stacked to form the FVTC correlation structure. FVTC includes every correlation within the considered D frames and also avoids the repetitive use of same correlations as in the TDEC correlation matrix.

\section{Multimodal systems}
\label{sec:multimodal}

\subsection{Parallel delay scale TDEC-CNN model (TDEC-CNN) :  Model 1}
\label{ssec:TDECCNN}
We developed a CNN architecture which takes in multiple time-delay embedded correlation matrices with 2 delay scales in parallel as inputs for two 2D-CNN layers. The output from the 2 CNN layers are then concatenated and passed through another 2D-CNN layer. Batch normalization, max-pooling and dropout were applied afterwards. The flattened output is then fed to a fully connected layer with 64 neurons. 16 filters with kernel size (3,3) was used for every 2D CNN layer and every CNN layer has ReLU activation. We trained individual models for FAUs, TVs and MFCCs where 3 and 7 sample delay scales were used for FAUs and 7 and 15 sample delay scales were used for TVs and MFCCs. 

\subsection{FVTC CNN model (FVTC-CNN) : Model 2}
\label{ssec:fvtcCNN}
We designed a CNN model inspired by the one in \cite{Huang2020ExploitingVT} which takes the FVTC correlation matrix computed in section \ref{ssec:FVTC} as the input. To reduce the number of trainable parameters in the original model \cite{Huang2020ExploitingVT}, we reduced the size of the two fully connected layers to 64 and 8. We used the same dilation rates {1,3,7,15} as in the original model. We chose 45 as the 'D' parameter for FAUs and 50 for TVs and MFCCs. The 'D' values were chosen from the set of (45,50,55) after doing a grid search on individual feature based systems.

\subsection{Multimodal CNNs}
\label{ssec:multimodalCNN}

\begin{figure*}[t]
  \centering
  \includegraphics[width=0.9\linewidth,height=34mm]{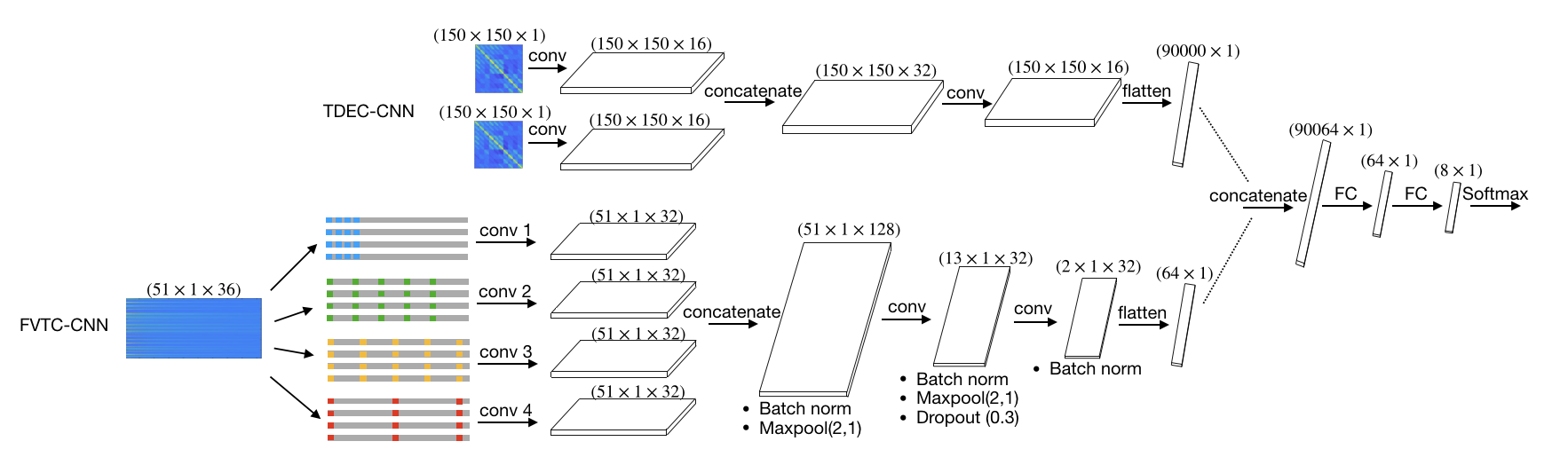}
  \caption{TDEC and FVTC combined multimodal architecture for best performing model in Table \ref{tab:modelresults2}}
    \label{fig:multimodel_architecture}
\end{figure*}

One of the key contributions of the paper comes with the mutimodal networks developed to fuse the audio and video based coordination features calculated from the TDEC and FVTC methods. 3 types of fusion models were designed namely the (Parallel delay TDEC-FVTC CNN), (FVTC-FVTC CNN) and (Parallel delay TDEC- Parallel delay TDEC CNN). The same model architectures developed in section \ref{ssec:TDECCNN} and section \ref{ssec:fvtcCNN} were used and they are fused after passing through all the 2D-CNN layers by concatenating the flattened outputs from each model. The concatenated output then passes through 2 fully connected layers which have 64 and 8 neurons, respectively. That output is then fed to a softmax output layer to generate the final class probabilities (for schizophrenic and healthy classes). 

To choose the learning rate and batch size for the individual and multimodal systems, we did a grid search from sets of (1e-4, 3e-4, 1e-5, 3e-5) and (32,64,128) for learning rates and batch sizes, respectively. 1e-4 for learning rate and 64 for batch size gave the best metrics in classification for the best performing model in Table \ref{tab:modelresults2}. Every model was trained in leave one subject out cross validation fashion where both accuracy and F1 scores are calculated across 18 folds. This ensured that every model is always trained with around 2000 sample points (17 subject's data) and then tested on the excluded subject (test fold). To come up with the subject level prediction label from the multiple segment level predictions, we use the best 25\% of the total segments in the test fold which are predicted with the highest class probabilities (even if the model predicts the wrong class). Every model is trained for 200 epochs with early stopping based on validation loss with patience of 15 epochs. Figure \ref{fig:multimodel_architecture} shows the architecture for the best performing multimodal system from Table \ref{tab:modelresults2}.

\begin{table}[t]
  \caption{Individual Model results for FAUs, TVs and MFCCs. Best Model for each feature type is highlighted in bold.}
  \centering
  \label{tab:Indmodresults}
  
    \begin{tabular}{p{25mm}p{9mm}p{8mm}p{9mm}p{8mm}}
    \toprule
    \textbf{Features} & \multicolumn{2}{p{25mm}}{\textbf{TDEC-CNN (Model1)}}    & \multicolumn{2}{p{23mm}}{\textbf{FVTC-CNN (Model2)}} \\
    \midrule
                          & Accuracy   & F1(S)/F1(H)     & Accuracy        & F1(S)/F1(H)     \\
FAU                       & \textbf{83.33\%} & \textbf{0.80/0.86} & 83.33\%          & 0.77/0.87          \\
TV                        & 66.67\%          & 0.57/0.73          & \textbf{72.22\%} & \textbf{0.62/0.78} \\
MFCC                      & 61.11\%          & 0.46/0.70          & \textbf{72.22\%} & \textbf{0.55/0.80} \\ 
MFCC+Glottal TVs                     &60.05 \%          &0.45/0.69          & \textbf{72.22\%} & \textbf{0.55/0.80} \\ 

\bottomrule
\end{tabular}
\label{tab:dataset}
\end{table}

\begin{table}[th]
\caption{Multimodal classification results}
\label{tab:modelresults2}
\small
\vspace{-5pt}
\centering
  
    \begin{tabular}{cp{10mm}p{10mm}p{12mm}c}
    \toprule
    \textbf{Models}                            & \textbf{Accuracy} & \textbf{F1(S)/F1(H)} \\ \hline
FAU (Model2)+TV(Model2)            & 66.67\%                & 0.67/0.67                          \\ 
FAU (Model1)+TV(Model1)              & 72.22\%                 & 0.67/0.76                          \\ 
FAU (Model2)+MFCC(Model2)        & 72.22\%                 & 0.62/0.78                          \\ 
FAU (Model1)+MFCC(Model2)        & 77.78\%                 & 0.67/0.83                  \\ 
FAU (Model1)+(MFCC+Glottal TVs)(Model2)        & 83.33\%                 & 0.73/0.88                  \\ 
\textbf{FAU (Model1)+TV(Model2)} & \textbf{88.89\%}        & \textbf{0.86/0.91}        \\ \hline

\end{tabular}
\end{table}


\section{Discussion}
\label{sec:typestyle}


The averaged eigenspectra and difference plots in figure \ref{fig:eigenspecs and cohends} shows that the low-rank eigenvalues are smaller for schizophrenic subjects relative to the healthy controls, and this trend is reversed towards the high-rank eigenvalues. A key observation associated with depression severity \cite{Williamson2014, WILLIAMSON2019, Espy-Wilson2019} is that low-rank eigenvalues are larger for MDD subjects relative to healthy controls where as they are smaller for high-rank eigenvalues. The magnitude of high-rank eigenvalues indicates the dimensionality of the time-delay embedded feature space. Thus, larger values in the high-rank eigenvalues can be associated with greater complexity of articulatory coordination\cite{Espy-Wilson2019, Williamson2013}. Therefore we can infer that the schizophrenic subjects with strong positive symptoms have a higher complexity than the healthy controls and the MDD patients. These results are likely due to the negative symptoms of depression which results in psychomotor slowing (i.e., simpler coordination) and the strong positive symptoms of the schizophrenic patients such as activation that results in motor hyperactivity (i.e., complex coordination).  Supporting our hypothesis, we see this effect from eigenvalues computed from the FAUs, TVs and MFCCs.


Table \ref{tab:Indmodresults} shows the average accuracy across the 18 folds and the F1 scores for classifying schizophrenic and healthy subjects by training individual models for FAUs, TVs and MFCCs based on TDEC-CNN and FVTC-CNN models. Results suggest that FAUs perform the best when compared to TVs and MFCCs in classification metrics.
This could be due to the inclusion of a wider range of facial muscle movements which are not limited to only those around the speech articulators. Moreover, FVTC-CNN models trained with TVs and MFCCs perform the best when compared to TDEC-CNN models trained with the same features. With respect to F1 scores, TVs perform better than the MFCCs in both FVTC-CNN and TDEC-CNN models showing the robustness of TVs in capturing the articulatory changes in speech.

Table \ref{tab:modelresults2} shows results for the 6 multimodal systems trained by fusing video and audio features from TDEC and FVTC methods. It is interesting to note that the models with heterogeneous architectures perform the best when compared to models which use the same correlation structure for both audio and video features. To do a fair comparison between TVs and MFCCs, we also trained individual and multimodal networks by incorporating the 2 glottal TVs along with the 12 MFCCs, so that glottal source level information is also accounted. Even then, Table \ref{tab:modelresults2} shows that the TV based best performing model out performs the MFCC based multimodal systems. Furthermore, the multimodal system which uses TDEC and FVTC correlation structures for FAUs and TVs, respectively, outperforms the baseline model trained on FVTC for both FAUs and MFCCs by around 18\% in terms of the mean F1 score. This result is interesting in the sense that the video features complement well with TVs over MFCCs in the proposed multimodal setting. 

In conclusion, this paper proposes a multimodal approach to classify subjects with strong positive symptoms in schizophrenia from healthy. We also show that the video based features are more effective in identifying articulatory coordination changes while also asserting the fact that fusing with audio based TVs significantly boost the performance in classification.  
\begin{acks}
This work was supported by a UMCP  UMB - AI + Medicine for High Impact (AIM-HI) Challenge Award.
\end{acks}

\bibliographystyle{ACM-Reference-Format}
\bibliography{mybib}


\end{document}